\documentclass[10pt,letterpaper,twocolumn]{article} 

\usepackage{ol2}
\usepackage[draft]{hyperref}
\usepackage{amsmath}

\begin{document}

\twocolumn[ 

\title{Experimental proposal for the generation of entangled photon triplets by third-order spontaneous parametric downconversion in optical fibers}


\author{Mar\'ia Corona,$^{1,2,*}$ Karina Garay-Palmett,$^1$ and Alfred B. U'Ren$^{1}$}

\address{
$^1$Instituto de Ciencias Nucleares, Universidad Nacional Aut\'onoma de M\'exico, apdo. postal 70-543, M\'exico 04510 DF
\\
$^2$Departamento de \'Optica, Centro de Investigaci\'on Cient\'{\i}fica y
de Educaci\'on Superior de Ensenada, Apartado Postal 2732, Ensenada,
BC 22860, M\'exico \\
$^*$Corresponding author: maria.corona@nucleares.unam.mx
}

\begin{abstract}
We present an experimental proposal for the generation of photon triplets based on third-order spontaneous parametric downconversion in thin optical fibers. Our analysis includes expressions for the quantum state which describes the photon triplets, and for the generation rate in terms of all experimental parameters.    We also present, for a specific source design, numerically-calculated generation rates.
\end{abstract}


 ] 

\bigskip

The generation of multi-photon entangled states represents an important goal in quantum optics, both for fundamental tests of quantum mechanics and for quantum-enhanced technologies.    Spontaneous parametric downconversion (SPDC) in second-order nonlinear crystals is the physical process of choice for most entangled photon-pair experiments.  Within the last decade, the process of spontaneous four wave mixing (SFWM) based on the third-order nonlinearity in optical fibers has emerged as a viable alternative to SPDC\cite{fiorentino02}. The same third-order nonlinearity which makes the SFWM process possible, could also serve as the basis for a different process: third-order spontaneous parametric downconversion (TOSPDC)\cite{chekhova05,Felbinger98,douady04}. While in the SFWM process two pump photons are annihilated in order to generate a photon pair, in the TOSPDC process a single pump photon is annihilated in order to generate a photon triplet.  In this paper we describe a proposal for the generation of photon triplets in thin optical fibers.

A number of approaches for the generation of photon triplets have been either proposed or demonstrated with extremely low collection efficiencies, including tri-excitonic decay in quantum dots~\cite{persson04}, combined second-order non linear processes~\cite{keller98}, and approximate photon triplets formed by SPDC photon pairs together with an attenuated coherent state~\cite{rarity98}.  The clearest demonstration of photon triplet emission, albeit with low count rates,  is a recent remarkable experiment involving two cascaded second-order SPDC processes~\cite{hubel10}. In contrast, our proposed technique permits the \textit{direct} generation of photon triplets, without postselection, and a straightforward source modification would lead to the direct generation of three-photon Greenberger-Horne-Zeilenger (GHZ) states\cite{greenberger90}.  Our proposal likewise permits experimental studies of the largely unexplored three-particle continuous variable (spectral) entanglement.

Borrowing from second-order SPDC terminology, we refer to the three photons in a given TOSPDC triplet as signal-$1$ (r), signal-$2$ (s) and idler (i).  Here we consider the TOSPDC process in an optical fiber, with the three generated photons in the same transverse mode, and where all four fields are co-polarized.  It can be shown~\cite{corona10} that in the spontaneous limit, the state of the emitted radiation can be written in terms of the three-photon component $|\Psi_3\rangle$ as

\begin{equation}
\label{quantState}|\Psi\rangle=|0\rangle_{r}|0\rangle_{s}|0\rangle_i+\zeta|\Psi_3\rangle,
\end{equation}

\begin{align}
\label{quantest3}|\Psi_3\rangle=&\sum_{k_{r}}\!\sum_{k_{s}}\!\sum_{k_i}G_k(k_{r},k_{s},k_i)\nonumber \\\times&\hat{a}^\dag(k_{r})\hat{a}^\dag(k_{s})\hat{a}^\dag(k_i)
|0\rangle_{r}|0\rangle_{s}|0\rangle_i,
\end{align}

\noindent written in terms of the wavenumbers $k_\mu$ for the four participating fields ($\mu=p,r,s,i$), the creation operator $\hat{a}^\dag(k)$, and the joint amplitude function $G_k(k_{r},k_{s},k_i)$. For a pulsed pump, centered at $\omega^o_p$, with bandwidth $\sigma$,  and with a Gaussian spectral envelope $\alpha(\omega)=\exp[-(\omega-\omega^o_p)^2/\sigma^2]$,  the quantity $\zeta$ (related to the conversion efficiency) is given by

\begin{equation}
\label{zeta} \zeta= \left[ \frac{2(2\pi)^{3/2}\epsilon_0^3c^3n_p^3PL^2\gamma^2}{\hbar^2(\omega^o_p)^2\sigma}\right]^{1/2}(\delta k)^{3/2}.
\end{equation}

$\zeta$ has been written in terms of the fiber length $L$, the pump peak power $P$, the refractive index $n_p \equiv n(\omega_p^o)$, the mode spacing $\delta k$, the vacuum electrical susceptibility $\epsilon_0$, and the nonlinear coefficient $\gamma$, defined as

\begin{equation}
\label{gamma}
\gamma=\frac{3\chi^{(3)}\omega_p^o}{4\epsilon_0c^2n_p^2A_{eff}},
\end{equation}

\noindent where $\chi^{(3)}$ is the third-order susceptibility and $A_{eff}=\left[\int\!\!dx\!\!\int\!\!dyf_p(x,y)f_{r}^*(x,y)f_{s}^*(x,y)f_i^*(x,y)\right]^{-1}$ is the effective interaction area, written in terms of the transverse distribution $f_\mu(x,y)$ for mode $\mu$.

Writing $G_k(k_{r},k_{s},k_i)$ in terms of frequencies leads to the joint spectral amplitude $G(\omega_{r},\omega_{s},\omega_i)= \ell(\omega_{r})\ell(\omega_{s})\ell(\omega_i)F(\omega_{r},\omega_{s},\omega_i)$, where $\ell(\omega)=\sqrt{(\hbar \omega)/( \pi \varepsilon_0 n^2(\omega))}$ ($n(\omega)$ is the index of refraction) and where the function $F(\omega_{r},\omega_{s},\omega_i)$ is given as the product of the pump spectral amplitude (PSA) function $\alpha(\omega)$ and the phasematching (PM) function $\phi(\omega_{r},\omega_{s},\omega_i)$, in turn given by

\begin{equation}
\label{JSA}\phi(\omega_{r},\omega_{s},\omega_i)=\mbox{sinc}\!\left[\frac{L}{2}\Delta k(\omega_{r},\omega_{s},\omega_i)\right]e^{i\frac{L}{2}\Delta k(\omega_{r},\omega_{s},\omega_i)},
\end{equation}

\noindent written in terms of the phasemismatch $\Delta k(\omega_{r},\omega_{s},\omega_i)=k_p(\omega_r+\omega_s+\omega_i)-k(\omega_r)-k(\omega_s)-k(\omega_i)+\Phi_{NL}$, which contains a nonlinear term $\Phi_{NL}=[\gamma_p-2(\gamma_{pr}+\gamma_{ps}+\gamma_{pi})]P$, expressed in terms of a self-phase modulation coefficient $\gamma_p$ and cross-phase modulation coefficients $\gamma_{p\mu}$.

From Eqs.~\ref{quantest3} and \ref{zeta} it can be shown~\cite{corona10} that the number of triplets generated per second, $N$, is given by

\begin{align}
\label{NumFot}N&= \lim_{\delta k \to\ 0} \sum_{k_r} \langle \Psi | a^\dag(k_{r})a(k_{r}) | \Psi \rangle R\nonumber \\
&=\frac{2^33^2\hbar c^3n_p^3}{\pi^2(\omega_p^o)^2}\frac{L^2\gamma^2 P R}{\sigma^2} \nonumber \\
&\times\!\! \int\!\!d\omega_{r}\!\!\int\!\!d\omega_{s}
\!\!\int\!\!d\omega_i\frac{k'_r\omega_r}{n_r^2}\frac{k'_s\omega_s}{n_s^2} \frac{k'_i\omega_i}{n_i^2} |F(\omega_{r},\omega_{s},\omega_i)|^2,
\end{align}

%

\noindent in terms of the pump repetition rate $R$, $k'_\mu \equiv k'(\omega_\mu)$,  and $n_\mu \equiv n(\omega_\mu)$ ($'$ denotes a frequency derivative).


The generation of frequency-degenerate photon triplets requires the fulfilment of $k(3 \omega)=3 k(\omega)$. In general, however, this is not trivial to attain; the large spectral separation between the pump and the generated photons implies that $k(3 \omega)$ is typically considerably larger than $3 k(\omega)$.   Our proposed solution exploits the use of two different transverse modes in a thin fiber guided by air, i.e. with a fused silica core and where the cladding is the air surrounding this core.  In particular, we will assume that while the TOSPDC photons all propagate in the fundamental mode of the fiber (HE$_{11}$), the pump mode propagates in the first excited mode (HE$_{12}$)\cite{grubsky07}.

\begin{figure}[t]
\centerline{\includegraphics[width=7.8cm]{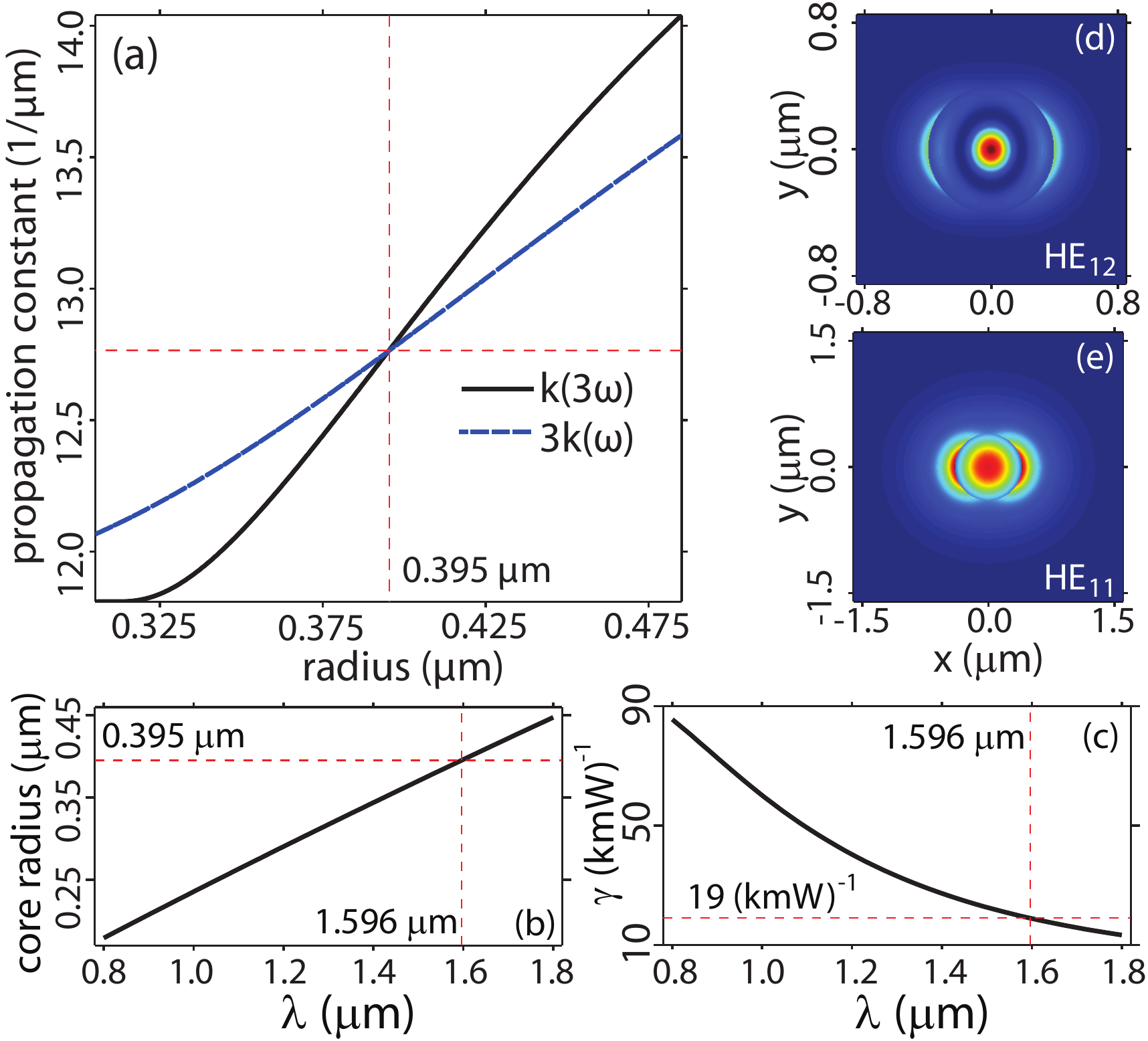}}
\caption{Frequency-degenerate phasematching for TOSPDC (a).  Phasematching radius (b) and nonlinearity $\gamma$ (c) vs degenerate TOSPDC wavelength.  Transverse intensity for the HE$_{12}$-mode pump (d) and for the HE$_{11}$-mode TOSPDC photons (e). }\label{PM}
\end{figure}

Figure~\ref{PM}(a) shows for a particular degenerate TOSPDC frequency $\omega$ (corresponding to $\lambda=1.596\mu$m), that a phasematching radius exists ($r=0.395\mu$m) for which $3 k(\omega)=k(3 \omega)$ is fulfilled. Note that for the pump powers to be considered here, $\Phi_{NL}$ can be neglected. We could choose a different $\omega$, which would lead to a behavior similar to that shown in figure~\ref{PM}(a), with a different resulting $r$;  the dependence of $r$ vs $\omega$ (expressed as wavelength) is shown in figure~\ref{PM}(b).   It turns out, as illustrated in Fig.~\ref{PM}(c), that a smaller $r$ (or smaller TOSPDC wavelength, according to the relationship shown in Fig.~\ref{PM}(b)) leads to a larger value for the nonlinearity $\gamma$.   Therefore, in order to obtain a large photon triplet flux (which scales as $\gamma^2$) it is advantageous to use small $r$ and consequently a large pump frequency $3 \omega$.   While this could suggest the use of an ultraviolet pump, in this paper we avoid the use of non-standard fiber-transmission frequencies.  Instead, we have chosen for our illustration in Fig.~\ref{PM}(a) a readily available pump frequency ($532$nm) resulting in photon triplets near the telecommunications band.   Note that while in general the required fiber radii are small, these can be obtained using current fiber taper technology, although with limited interaction lengths.

Figure~\ref{PM}(d) shows the pump transverse intensity, in mode HE$_{12}$, evaluated at $\lambda_p=1.596/3=0.532\mu$m.  Figure~\ref{PM}(e) shows the transverse intensity for the generated photons, in mode HE$_{11}$, evaluated at $\lambda=1.596\mu$m. For this wavelength combination we obtain from Eq.~\ref{gamma} that $\gamma=19$(kmW)$^{-1}$. Starting from, say, a Gaussian pump beam in free space in general only a fraction of the power will be coupled to the HE$_{12}$ fiber mode; for the fiber assumed here, a beam radius of $r=0.783\mu$m (with the beamwaist at the air-crystal interface) exhibits a maximum overlap of $29.8\%$ with the HE$_{12}$ mode amplitude.

\begin{figure}[t]
\centerline{\includegraphics[width=8.3cm]{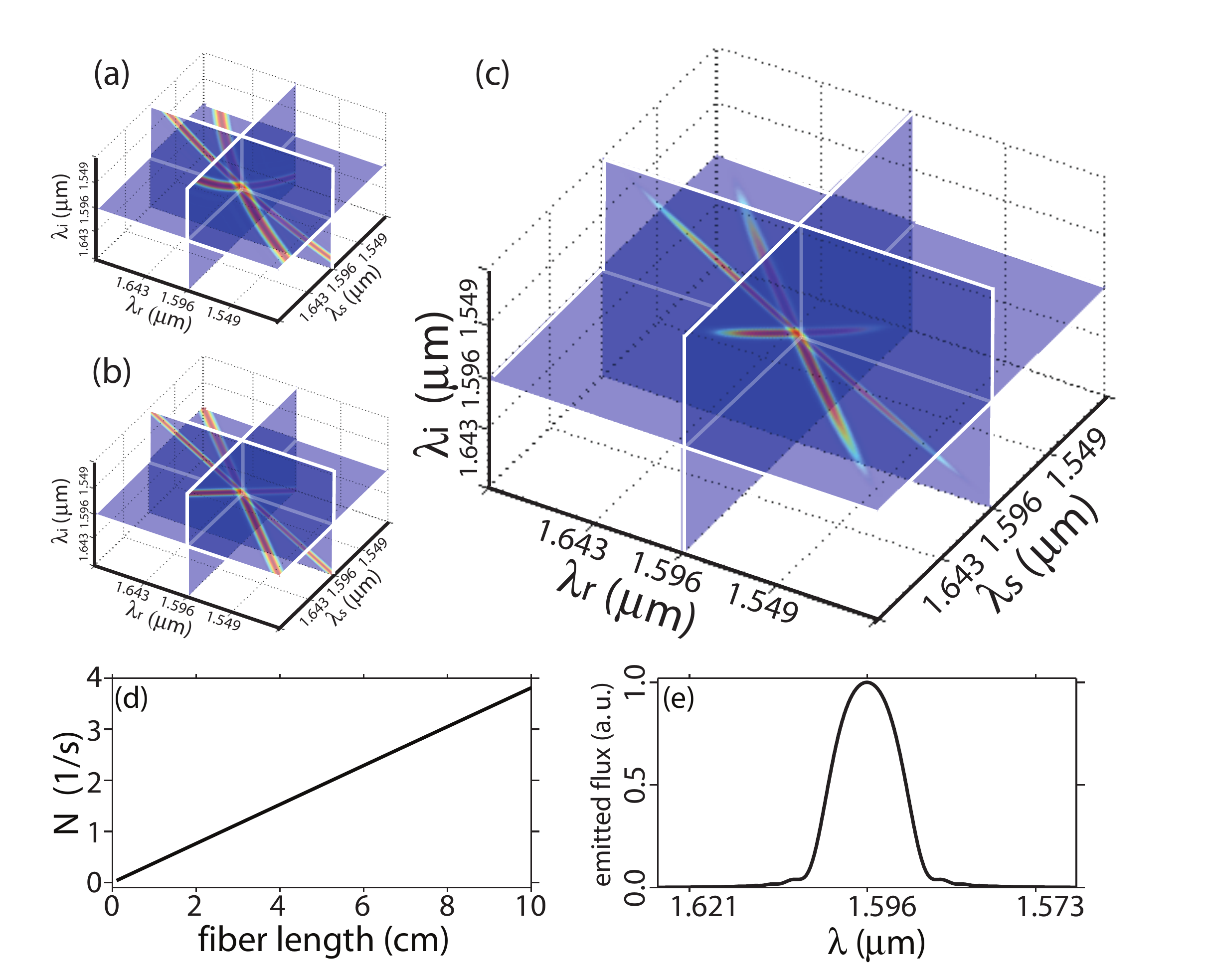}}
\caption{Representation, plotted as a function of the three generation frequencies, of: (a) phase matching function, (b) pump envelope function and (c) joint spectral intensity. (d) Number of photon triplets generated per second vs fiber length. (e) TOSPDC single-photon spectrum.}\label{JSI}
\end{figure}

Figure~\ref{JSI} shows a representation of the three-photon joint spectral intensity plotted as a function of the TOSPDC frequencies.  Figure~\ref{JSI}(a) shows the PM function, fig.~\ref{JSI}(b) shows the PSA function, while fig.~\ref{JSI}(c) shows the joint intensity.  Note that the tilted orientation of the joint intensity, resulting from a narrow width along the direction $\omega_{s1}+\omega_{s2}+\omega_{i}$ and much larger widths in the two transverse directions, indicates the presence of spectral correlations amongst the three photons, which underlie the existence of three-partite entanglement. Note that these plots are clear generalizations of similar plots for second order SPDC~\cite{grice97}.   We note that the values for $L$ and $\sigma$ (which control the widths of the PM and PSA functions respectively) assumed in our source design, presented below, lead to acute spectral correlations which are difficult to plot clearly.  Thus, in Fig~\ref{JSI}(a)-(c) we have assumed $L=0.6\mu$m and $\sigma=5.1$THz, which are considerably smaller and larger, respectively, compared to values assumed for our source design, but which lead to the same qualitative behavior.

Evidently, a crucial consideration is the attainable source brightness, which we derive from numerical integration of Eq.~\ref{NumFot};  we assume $\sigma=23.5$GHz (which corresponds to a pulse duration of $100$ps), an average pump power of $200$mW and $R=100$MHz. Figure~\ref{JSI}(d) shows the number of generated photon triplets per second plotted as a function of the fiber length, exhibiting the expected linear dependence. We have limited the fiber length to $L=10$cm for which we obtain a generation rate of $3.8$ triplets per second; note that based on recent publications, it is possible to obtain uniform-radius tapers with $\sim 445$nm radius over a $9$cm length\cite{lsaval04}.  Fig.~\ref{JSI}(e) shows for a fiber length of $L=10$cm, the resulting single-photon spectrum for any of the three emission modes.  Note that because the three photons in a given triplet are in the same spatial and spectral mode, triplet splitting can only be attained probabilistically; in Ref.~\cite{corona10} we discuss spectrally non-degenerate TOSPDC.

There are some important differences between the TOPDC and SFWM processes. In common with second-order SPDC, the emitted flux in TOSPDC scales linearly with pump power and is constant (within the phasematching bandwidth) with respect to the pump bandwidth. In contrast, for SFWM the emitted flux scales quadratically with pump power and linearly with pump bandwidth~\cite{garay10}.   Thus, SFWM sources tend to be significantly brighter than TOSPDC sources.  Nevertheless, the source studied here leads to $3$ orders of magnitude greater flux than that reported in \cite{hubel10}.  Another important difference is that while for many SFWM designs contamination due to spontaneous Raman scattering is a concern (signal and idler modes are often within the Raman gain spectral window), the inherent large spectral separation between the pump and the generated photons in the case of frequency-degenerate TOSPDC, eliminates this concern.

An additional possible source of noise arises if pump spectral broadening due to self phase modulation is sufficient to encompass the TOSPDC emission modes.   The large spectral separation between pump and TOSPDC photons helps in this regard; additionally, this effect may be controlled by restricting the pump bandwidth and power.   Pulse propagation simulations (not shown) indicate that contamination due to this mechanism is not a concern for the source design presented here.

Note that if the fiber used for generating photon triplets is pumped in both directions, through the use of a Sagnac interferometer, it becomes possible to generate three-photon GHZ states of the type $2^{-1/2}(|HHH \rangle + |VVV \rangle)$ ($H$/$V$ represent horizontal/vertical polarization) through a modified version of the source proposed here. This is similar to a scheme used for generating polarization entangled photon pairs through SFWM~\cite{lee06}.

We have presented an experimental proposal for the generation of photon triplets based on third-order spontaneous parametric downconversion in thin optical fibers.  We have shown that phasematching can be attained if the pump propagates in the mode HE$_{12}$ while the generated photons propagate in the mode HE$_{11}$. We have presented an expression for the joint amplitude describing the photon triplets, as well for the expected source brightness.    We have presented a particular source design together with the expected generation rates. We expect that the technique presented will be useful for the exploration of multi-photon entangled states.

This work was supported in part by CONACYT, Mexico,  by DGAPA, UNAM and by FONCICYT project 94142.

\end{document}